\documentclass[twocolumn,english,aps,pre,showpacs]{revtex4}
\usepackage[T1]{fontenc}
\usepackage[latin1]{inputenc}
\usepackage{float}
\usepackage{amsmath}
\usepackage{graphicx}
\usepackage{amssymb}

\makeatletter

\newcommand{\Ci}[1]{ }
\newcommand{\Cf}[1]{ }

\usepackage{babel}
\makeatother
\begin{document}

\preprint{preprint}

\title{Quantum phase transition in the Frenkel-Kontorova chain:\\
from pinned instanton glass to sliding phonon gas}

\author{O.V. Zhirov}
\email{zhirov@inp.nsk.su}
\affiliation{Budker Institute of Nuclear Physics, 630090 Novosibirsk, Russia}

\author{G. Casati}
\email{giulio.casati@uninsubria.it}
\affiliation{International Center for the Study of Dynamical 
Systems, Universit\`a degli Studi dell'Insubria and\\
Istituto Nazionale per la Fisica della Materia, 
Unit\`a di Como, Via Valleggio 11, 22100 Como, Italy and\\
Istituto Nazionale di Fisica Nucleare, 
Sezione di Milano, Via Celoria 16, 20133 Milano, Italy}

\author{D.L. Shepelyansky}
\email{dima@irsamc.ups-tlse.fr}
\affiliation{Laboratoire de Physique Quantique, UMR 5626 du CNRS, Universit\'{e}
Paul Sabatier, 31062 Toulouse, France}

\date{October 17, 2002}

\begin{abstract}
We study analytically and numerically the one-dimensional quantum Frenkel-Kontorova
chain in the regime when the classical model is located in the pinned 
phase characterized by the gaped phonon excitations and devil's staircase.
By extensive quantum Monte Carlo simulations we show 
that for the effective  Planck constant $\hbar$  
smaller than the critical value
$\hbar_c$ the quantum chain is in the pinned instanton glass phase.
In this phase
the elementary excitations have two branches:
\emph{phonons}, separated from zero energy by a finite gap, and \emph{instantons} which
have an exponentially small excitation energy. At  $\hbar=\hbar_c$
the quantum phase transition
takes place and for $\hbar>\hbar_c$ the pinned instanton glass is transformed into the
sliding phonon gas with gapless phonon excitations.
This transition is accompanied by the divergence of the spatial correlation
length and appearence of sliding modes at $\hbar>\hbar_c$.
\end{abstract}
\maketitle

\section{introduction}

The Frenkel-Kontorova (FK) model \cite{FK38} describes a one-dimensional
chain of atoms/particles with harmonic couplings placed in a periodic
potential. This model was introduced more than sixty years ago with
the aim to study crystal dislocations \cite{FK38,Naba67}. It was
also successfully applied later to the description of commensurate-incommensurate
phase transitions \cite{VPaAT84}, epitaxial monolayers on the crystal
surface \cite{SY71}, ionic conductors and glassy materials \cite{Piet81,Au78,Au83a}
and, more recently, to charge-density waves \cite{Flor96} and dry
friction \cite{Brau97,Cons00}. Despite the fact that the relevant
phenomena are at the atomic scale, all these works are based essentially
on the classical approach. The first study of quantum effects was done 
twelve years ago \cite{Borg89,Borg90,Borg89d} with the attempt
to understand the highly nontrivial quantum ground state of the Frenkel-Kontorova model
in the regime when the classical ground state is characterized
by the fractal ``devil staircase'' \cite{Au78}. These studies were
extended in 
\cite{Berm94,Berm97} and later the quantum dynamics at different values
of quantum parameter $\hbar $ was studied in \cite{Hu98,Hu99,Hu01,Hu01r,Ho01}.

The physical properties even of the classical FK model are very rich 
and nontrivial. In 1978 Aubry discovered \cite{Au78} a new type of ground
state which has fractal properties known as ``devil's staircase''. 
In fact, the equilibrium positions of atoms in the FK chain 
are described \cite{Au78} by the well known Chirikov standard map \cite{Chir79},
which describes generic properties of chaotic Hamiltonian dynamics.
The density of particles in the equilibrium
state determines the rotation number of the invariant curves of the
map, while the amplitude of the periodic potential in the FK model
gives the value of the dimensionless parameter $K$. According to
the known properties of the map, it follows that there exist two phases
of the chain, the ``sliding'' phase and the ``pinned''
phase. Indeed at $K<K_{c}$ the Kolmogorov-Arnold-Moser (KAM) curves are smooth and 
the chain can easily slide along the potential. This implies the existence of
a zero phonon mode. In contrast, at $K>K_{c}$ the KAM curves are
destroyed and replaced by an invariant Cantor set which is called
cantorus \cite{Au78,Perc79,Au83b,Au83c,Mac84}. In this ``pinned'' phase the chain cannot
slide being kept by a finite Peierls-Nabarro barrier and the phonon
spectrum is separated from zero by a finite gap. In this paper we consider only
the gaped/pinned phase with  $K>K_{c}$.

It is important to stress that in the pinned phase, besides the equilibrium state
of minimal energy, there exists a lot of other \emph{equilibrium} configurations,
corresponding to \emph{local} minima of the potential with energies
very close to the minimal energy, i.e. the energy of the ground state
\cite{Sch84,Sch85,Zhir01a}. The total amount of these states (configurational
excitations) grows exponentially both with the length $L$ of the
chain, as well as with the parameter $K$ \cite{Zhir01a}. Moreover, a great number of
them is practically degenerate since their energy separation from
the ground state is exponentially small. 

In the classical limit all these configurational excitations
are stable, while in the quantum case they become metastable due to tunneling
between different exponentially degenerate minima of the potential. As 
a result one may expect that the true eigenstates of the Hamiltonian, including
the ground state, are built as superpositions of classical configurational
states. Such a nontrivial structure of the quantum ground state of the chain has
a profound  analogy with the famous vacuum of Quantum
Chromodynamics (QCD) \cite{Shur98} in which tunneling transitions between different,
practically degenerate, states are known as ``instantons''.
This analogy between the two problems is very useful and implies 
that many of the methods developed in lattice QCD studies may
be applicable to quantum FK model.

Dynamical low energy excitations in the classical FK chain consist
only of phonon modes describing small vibrations around a classical minimum
of the chain potential energy. In the quasiclassical regime, where  
the effective dimensionless Planck constant $\hbar$ is very small,
the quantization of the FK chain can be reduced to a quantization of phonon
modes only. Indeed the time of tunneling between different minima of
the potential energy is exponentially large, compared to periods of
the vibrations, and the phonon modes are decoupled from the tunneling
modes (instantons). 
This reminds the situation in the QCD \cite{Shur98}
where the quasiclassical regime for instantons appears at small distances
on which instantons are also decoupled from other excitations like 
quarks and gluons.

In this regime the tunneling transitions are very slow, instantons are local
and frozen in space. Due to exponential degeneracy of chain configurations the
model reveals a glass-like structure of instantons randomly distributed   
along the chain. We will call this phase the ``instanton glass''.

One may expect that the phase structure changes significantly with the increase 
of $\hbar$ when the
tunneling time $t\sim \exp (S_{in}/\hbar )$ ($S_{in}$ is the instanton
action) becomes comparable to the inverse frequency of phonons. In this
case phonon and instanton excitations can strongly influence each other
due to anharmonicity of the periodical
potential. This may lead to quantum melting of the instanton glass phase
and transition to another phase which appears above some critical value 
$\hbar>\hbar_c$.

The existence of another quantum phase can be argued in the following way.  
At sufficiently large  $\hbar \geq \hbar _{c}$ the kinetic energy of
quantum particle $E_{k}\sim \hbar^{2}/2m(\Delta x)^{2}$ starts to exceed
the height $U$ of the periodic potential in which the chain is placed
(here $\Delta x$ is the mean particle separation in the chain which is comparable
with the period of the potential, $m$ is the particle mass). 
The condition $E_{k}\sim U$ gives $\hbar_c\sim\Delta x\sqrt{mU}$.
As a result, for $\hbar>\hbar_c$ the chain turns from the \emph{pinned} 
to the \emph{sliding} phase. In this regime tunneling is replaced by direct 
propagation above the barrier.
Hence, the instanton-like motion is replaced by phonon-like motion corresponding 
to a new phase with gapless phonon excitations. This regime can be considered as 
sliding quantum phase similar to classical sliding regime at $K<K_c$ \cite{Au83b,Au83c}.
Qualitative different types of the behavior of quantum FK model at small and large
$\hbar$ values were already seen in the first numerical studies \cite{Borg89,Borg90}.  
In summary, at $\hbar =\hbar _{c}$
one may expect a \emph{quantum phase transition}, with qualitative
rearrangement of the spectrum of elementary quasiparticle excitations.

In this work, devoted to a comprehensive numerical study of the transition
we have just outlined, we present a complete picture of low energy
excitations in the quantum FK chain. We also find that this
transition is highly nontrivial and the model reveals a non-analytical
behavior which can partially explain the failure of simple analytical
approaches developed in \cite{Berm94,Hu98,Hu99,Ho01}. It should be
stressed also that our results have relevance to a wider class of
quantum systems, like e.g. quantum spin glass or other disordered systems 
with interactions. Indeed, the existence of highly degenerate classically 
stable configurations is a general property of such systems.

The paper is organized as follows. In Section \ref{sec:the-fk-model}
we outline the model, its quantum features and the main points of our
numerical approach. In Section \ref{sec:elementary-excitations.}
we study  elementary excitations of the chain at different
values of $\hbar $ and show that there exists a structural rearrangement of
the excitation spectrum at certain $\hbar = \hbar_c$. 
A detailed analysis of this rearrangement, given in Section \ref{sec:QFT}, 
indicates that we have a quantum phase transition. Our results are 
summarized in Section \ref{sec:Conclusions}.

\section{The quantum Frenkel-Kontorova model}
\label{sec:the-fk-model}

\subsection{Definitions and outline of quantum features}

The model describes a one-dimensional chain of particles with harmonic
couplings placed in a periodic potential. The Hamiltonian reads:
\begin{equation}
  H=\sum _{i=0}^{s}\left[\frac{P_{i}^{2}}{2m}+\alpha\frac{(x_{i}-x_{i-1})^{2}}{2}-
  \beta\cos(x_{i}/d)\right]
\label{H_FKorig}
\end{equation}
 where $s$ is the number of particles, $P_{i}$ and $x_{i}$ are
their momenta and coordinates, and in the quantum case 
$P_{i}=-i\hbar\partial/\partial x_i$. In this paper we use  
units $m=d=\alpha=1$, that correspond to dimensionless Hamiltonian 
(see for details e.g. \cite{Borg89,Borg90,Berm94}):
\begin{equation}
  H=\sum _{i=0}^{s}\left[\frac{P_{i}^{2}}{2}+\frac{(x_{i}-x_{i-1})^{2}}{2}-
  K\cos(x_{i})\right]
\label{H_FK}
\end{equation}
with a dimensionless parameter $K=\beta/\alpha d^2$ and the
dimensionless Planck constant $\hbar$ is measured in units of $d^2\sqrt{m \alpha}$. 
In this way, $K$ is the chaos parameter in the Chirikov standard map \cite{Chir79,Au83b}.
We use the standard boundary conditions
\begin{equation}
  x_{0}=0,\quad x_{s}=L\, ,\label{Bcond}
\end{equation}
where the chain length $L=2\pi \cdot r$ consists of $r$ periods/wells 
of the external field. We analyze the standard case of golden mean ratio corresponding 
to $r/s\to(\sqrt{5}-1)/2$ (see, \cite{Borg89,Borg90,Zhir01a}).
The potential energy of the chain
\begin{equation}
  U(\{x\})=\sum _{i=0}^{s}\left[\frac{(x_{i}-x_{i-1})^{2}}{2}-
  K\cos (x_{i})\right]
\label{Upot}
\end{equation}
has a large number of minima, corresponding to different possible
distributions of particles among the wells. The classical 
ground state (absolute minimum of the potential energy) is
characterized by some special ordering discovered by Aubry 
\cite{Au78,Au83b,Au83c}. In addition, there are also local
minima, known as ``configurational excitation'' states\cite{Be80}.
Many of them have exponentially small energy separations from
the energy of the classical ground state \cite{Zhir01a}. In the classical
case all these states are well defined (distinguishable) and absolutely
stable.

In contrast, in the \emph{quantum} world these states are \emph{metastable}
due to quantum tunneling. Since they are practically degenerate with
the classical ground state the actual \emph{quantum} ground state
is built by \emph{many} of them. The admixture of metastable classical 
configurations in the quantum ground state was first discussed in \cite{Berm94,Berm97}. 
Let us summarize below 
the most important aspects of the quantum ground state, related to
the tunneling between these configurations.

In the quasiclassical region, the transition amplitudes between different
metastable configurations are exponentially small. Hence,
the ground state wave function is (with exponential accuracy) a sum
of \emph{non}-overlapping parts, each referring to a particular classical
configuration. Any average over the quantum ground state is (within
the same accuracy) a weighted sum of averages over relevant classical
configurations. Note that in this limit the main contribution to quantum
motion comes from phonons which characterize small vibrations around a 
classical equilibrium 
configuration. In fact, the phonon spectrum is only weakly sensitive
to the choice of specific configuration \cite{PhonFK} and 
the influence of tunneling processes on the global (thermodynamical) 
properties of the chain is expected to be small. In particular,
the phonon spectrum is still characterized by a phonon gap similar to 
the classical case. In this regime the tunneling 
transitions between metastable configurations are very slow comparing to 
phonon frequencies and can be considered as well separated instantons.

At higher values of $\hbar $ the tunneling rate increases and becomes
comparable with the frequency of phonons. In this regime phonon
oscillations are large and essentially anharmonic and therefore the
known analytical approaches \cite{Berm94,Hu98,Hu01r,Ho01} based on
the gauss-like wave function profile are not quite adequate. Instead, 
interactions between instantons and phonons come into play here.
As it will be shown later,  this leads to a new sliding phase appearing
at $\hbar>\hbar_c$. 

In fact, the first effects of quantum tunneling between classical configurations
were already seen in the numerical studies \cite{Borg90} (see Fig.4b). However, 
they were not attentively analyzed and the mixture of classical metastable 
configurations induced by quantum tunneling was not discussed. 
Certainly, tunneling processes affect significantly the average positions of 
particles at large time scales. Therefore the quantities based on mean expectation 
values of particle positions are not adequate even in the deep quasiclassical regime. 
Hence the basic concepts of classical treatment, like hull \cite{Au78,Au83b}
and g-functions \cite{Borg89,Borg90}, are also not adequate in this quantum regime.
The point is that in the case of particle tunneling between
two classical equilibrium positions the expectation value
does not correspond to any \emph{probable} (in the classical sense)
particle position. This is evident from the analogy with the two-well
potential problem, where the mean expectation value of a particle
position coincides with the \emph{top} of the barrier.

\subsection{Path integral and numerical simulations}
\label{sub:Path-Sim}

Tunneling effects are best understood in the Feynman path integral
formulation of quantum mechanics\cite{Poly77}. In particular, the
transformation to ``Euclidean'' time variable $\tau$: $t=i\tau$
makes the tunneling transition, or \emph{instanton,} local in time
$\tau$. In the quasiclassical regime the tunneling probability
is very small and therefore the mean separation between instantons 
is large compared to their size. This allows to use the approximation
of dilute instanton gas \cite{Poly77}. 

Feynman path formulation in the ``Eucleudean'' time 
allows direct numerical simulations of quantum systems with
probabilistic treatment \cite{Creu81} of the path integral:
\begin{equation}
  Z=\int Dx[\tau ]\exp (-\frac{1}{\hbar }S[x(\tau )]),
\label{FeyP}
\end{equation}
where the ``Eucledian'' action
\begin{equation}
  S[x(\tau )]=\int _{0}^{\tau _{0}}d\tau \sum_i
  \left(\frac{\dot{x}_{i}^{2}}{2}+\frac{(x_{i}-x_{i-1})^{2}}{2}-
  K\cos (x_{i})\right)
\label{Sc}
\end{equation}
has the same form as a total energy of a chain of particles in the usual 
real time variable, integrated over some time interval $\left[0,\tau _{0}\right]$.
This is equivalent to the consideration of the quantum system at some finite
temperature\cite{Creu81}
\begin{equation}
  T=\hbar /\tau _{0}  %\label%{eq:Tempr}
\label{eq:Temperature} 
\end{equation}

We assume periodic boundary conditions in the \emph{time}
direction, which correspond to a path closed on a torus 
\begin{equation}
  x_{i}(\tau _{0})=x_{i}(0),\quad i=1,\ldots ,s-1,
\label{TbndCnd}
\end{equation}
and integrate over the initial conditions $x_{i}(0)$ in order to
restore homogeneity of paths along the time torus. This allows to
improve the statistics in data measurements by averaging the data
along the torus.

The numerical simulation of the path integral (\ref{FeyP}) needs
discretization of the time variable $\tau _{n}=\Delta \tau \cdot n$,
$n=1,\ldots ,N$ by splitting the time interval $\tau_ {0}$ into $N$
steps of size $\Delta \tau =\tau_0 /N$. As a result the original 
time-continuous model (\ref{H_FK}) turns into a 2-dimensional lattice 
model with the action
\begin{eqnarray}
  S & = & \sum _{n=0}^{N}\sum _{i=1}^{s}\left[\right.
        \frac{(x_{i,n+1}-x_{i,n})^{2}}{2\Delta \tau }+\nonumber \\
    &   & +\Delta \tau \frac{(x_{i+1,n}-x_{i,n})^{2}}{2}-
        \Delta \tau \cdot K\cos x_{i,n}\left.\right].
\label{eq:S_Lat}
\end{eqnarray}

The time step $\Delta \tau $ should be chosen small enough to approach
the continuous limit of the original model (\ref{H_FK}). This leads
to the following requirements. At any time step $\Delta \tau $
the path variable $x_{i}$ jumps by a random shift 
$\Delta x_{i}\sim \sqrt{\hbar \Delta \tau }$. The first obvious requirement 
is that the shift $\Delta x_{i}$ should be small compared to the spatial scale 
of the potential energy variation.
Another requirement comes from the standard derivation of the path
integral \cite{Creu81,Shur84}: the potential energy terms should
be small compared to the kinetic energy:
\begin{equation}
  \Delta \tau \cdot \left(\frac{(x_{i+1,n}-x_{i,n})^{2}}{2}-K\cos x_{i,n}\right)
  \ll \frac{(x_{i,n+1}-x_{i,n})^{2}}{2\Delta \tau }.
\label{P<<K}
\end{equation}
The latter condition makes our lattice \emph{anisotropic} with respect
to spatial/time directions. In spite of the common belief \cite{cQFT97}
on the equivalence between quantum 1-dimensional chains and classical
2-dimensional statistical (e.g. \textsl{XY}) models, the above discussion
shows that some particular care has to be taken.

An accurate treatment of particle tunneling between two wells in the chain
requires some modifications of the action (\ref{eq:S_Lat}).
For a finite (not very small!) time step $\Delta \tau $, there exists
a probability for a particle to jump over the potential barrier in
\emph{one} time step\cite{Shur84}. The corresponding path contribution
to the action will be strongly underestimated by eq.(\ref{eq:S_Lat})
and, respectively, the probability of tunneling will be too high.
To cure this problem we use in our simulation an improved \cite{Shur84}
version of action (\ref{eq:S_Lat}), with the potential energy term
$\Delta \tau \cdot U(x_{n})$ replaced by the integral 
$\Delta \tau \cdot \int _{x_{n}}^{x_{n+1}}U(x)dx/(x_{n+1}-x_{n})$
along a straight line that links subsequent (in time) points $(x_{n},x_{n+1})$
of the particle path. This significantly improves the accuracy of 
numerical simulations.

For simulations of path ensembles we use the standard Metropolis algorithm
\cite{Metr53}. Each iteration looks as follows: at any fixed time
slice at number $n$ we update sequentially the particles coordinates
$x_{i,n}$, $i=1,\ldots ,s$; then we go to the next time slice: $n\rightarrow n+1,$
and so on. 

The system has two principal characteristic relaxation time scales
that are originated by different underlying processes. Basically one
can estimate the number of iterations as $N_{it}\sim 1/\omega ^{(min)}\Delta \tau $,
where $\omega^{(min)}$ is the lowest frequency relevant
to the process and $\Delta \tau $ is time discretization step. The
shortest scale is related to the path relaxation with respect
to main phonon modes, and is typically of the order of a few tens of iterations
in the quasiclassical regime, where phonons have a gap of
order unity. Another time scale is much larger and is determined by the smallest
frequency related to either the tunneling rate between
different classical configurations, or to the lowest phonon
frequency available in the system for $\hbar \geq \hbar _{c}$.

For the initial state we choose the Aubry classical ground state. Then, applying 
iterations, we generate a path ensemble. To be sure that the system does actually 
relax to statistical equilibrium with respect to the slowest processes described above, 
we control the mean number of particle path crosses a top of the potential 
barrier and we discard all configurations in the ensemble until this quantity
stabilizes. All computations are done for the chaos parameter $K=5$.
The required number of iterations to reach relaxation is very sensitive 
to the value of quantum parameter $\hbar $: for example at $\hbar =3$ this number is
$N_{it}\sim (2\div 4)10^{2}$ , while  at $\hbar=1$ it is  of the order of $10^{5\div 6}$.
This explains why in the first studies \cite{Borg89,Borg90} done at
$N_{it}\leq10^4$ many details at $\hbar \leq 2$ were not seen.
We study chains with up to 233 particles.

\section{elementary excitations}
\label{sec:elementary-excitations.}

The most important information about the quantum system is contained
in its spectrum of low-lying elementary excitations. Being the net
manifestation of system internal structure it reflects any structural
transition which can occur in the system, and it provides a complete
description of low-temperature thermodynamic and kinetic
properties. We extract this spectrum using a novel approach based
on the analysis of Fourier spectrum of Feynman paths. This approach
provides a most direct way to see and resolve different excitations
in the system. Then  we compare our method with the more traditional
one, based on the study of time correlation functions. This comparison 
provides a self-consistency check and demonstrates the advantages 
of our method.

\subsection{Spectral properties of Feynman paths}
\label{sub:SpFunc}

In classical nonlinear dynamics, the Fourier analysis of trajectories plays
a key role in understanding of periodic motion of complex systems. In
a similar way,
the spectral characteristics of Feynman paths are closely related
to the properties of elementary excitations in quantum systems. We start
our studies from the quasiclassical limit $\hbar \to 0$, where
this relation is exact, and extend them to higher values of
$\hbar $.

Let us consider the Fourier image of the path variable $x_{i}(\tau )$:
\begin{equation}
  a_{i}(\omega _{m}) = \frac{1}{\sqrt{\tau _{0}}}\int _{0}^{\tau _{0}}d\tau 
  \, x_{i}(\tau )\, \exp (i\omega _{m}\tau )
\label{eq:four_x}
\end{equation}
where $\omega _{m}=m\varpi $, $\varpi \equiv 2\pi /\tau _{0}$, and
$-\infty <m<\infty $. The path variable $x_{i}(\tau )$ is real,
therefore $a_{i}(-\omega _{m})=(a_{i}(\omega _{m}))^{\ast }$. To
get insight into the physical content of this quantity let us consider
the quasiclassical regime $\hbar \ll 1$. Then, for small variations
$x_{i}(\tau )=\overline{x}_{i}+\delta x_{i}(\tau )$ around the classical
static trajectory $\{\overline{x}_{i}\}$, one can expand the action
(\ref{Sc}) up to second order terms in $\delta x_{i}(\tau )$. Next,
using the spectral expansion for $\delta x_{i}(\tau )$ and performing
the integration over $\tau $, one gets:
\begin{eqnarray}
 S & = & S_{0}[\overline{x}]+\nonumber \\
   & + & \sum _{m}\sum _{i,k}\frac{1}{2}(\omega _{m}^{2}\delta _{ik}-
   \Omega _{ik}^{2})a_{i}(-\omega _{m})a_{k}(\omega _{m})\label{eq:Sexpa}\\
   \Omega _{ik}^{2} & = & \delta _{ik}\left(2+K\cos (x_{i})\right)-
   \delta _{i,k-1}-\delta _{i-1,k}\label{eq:Om2}
\end{eqnarray}
Now, by the transformation to normal modes 
$A^{(l)}(\omega _{m})=\sum _{i}V_{i}^{(l)}a_{i}(\omega _{m})$,
where eigenvectors $V_{i}^{(l)}$ satisfy the equation 
$\Omega _{ik}^{2}V_{k}^{(l)}=\nu _{l}^{2}V_{i}^{(l)}$
one gets the standard representation of the action as a sum of independent
phonon modes:
\begin{equation}\label{eq:Om3}
S=S_{0}[\overline{x}]+\sum _{m}\sum _{l}\frac{1}{2}(\omega _{m}^{2}+\nu _{l}^{2})\left|A^{(l)}(\omega _{m})\right|^{2}
\end{equation}
Finally, the path integral (\ref{FeyP}) turns into a product of ordinary
integrals
\begin{eqnarray*}
Z & = & \int \prod _{l=1}^{s-1}dA^{(l)}(0)\exp \left(-\nu _{l}^{2}\left|A^{(l)}(0)\right|^{2}/2\hbar \right)\times \\
 &  & \quad \prod _{m=1}^{\infty }d{\textrm{Re}}A^{(l)}(\omega _{m})d{\textrm{Im}}A^{(l)}(\omega _{m})\times \\
 &  & \quad \exp \left(-(\omega _{m}^{2}+\nu _{l}^{2})\left|A^{(l)}(\omega _{m})\right|^{2}/\hbar \right),
\end{eqnarray*}
and one arrives to the well known result for the correlator of free phonon
modes: 
\begin{equation}
 \left\langle A^{(l)}(\omega _{m})A^{(l)\ast }(\omega _{m'})\right\rangle =
 \frac{\hbar \delta _{mm'}}{(\omega _{m}^{2}+\nu _{l}^{2})}.
\label{eq:corrAA}
\end{equation}
We note that this result is obtained from Eq.(\ref{Sc}) where the
action is \emph{continuous} in time variable. In the discretized version
(\ref{eq:S_Lat}) the number of harmonics is finite: $\left|m\right|=0,1,\ldots ,M$,
where $M=\tau _{0}/2\Delta \tau $. It can be shown that the only modification 
induced by discretization is the replacement in (\ref{eq:corrAA}):
$\omega _{m}\rightarrow \widetilde{\omega }_{m}=(2\omega _{M}/\pi )\sin (\pi \omega _{m}/2\omega _{M})$,
$\omega _{M}\equiv M\varpi $. As a result, the spectral function 
for phonons is given by
\begin{equation}
 F^{(l)}(\widetilde{\omega }_{m})\equiv 
   \left\langle \left|A^{(l)}(\omega _{m})\right|^{2}\right\rangle =
   \frac{\hbar }{(\widetilde{\omega }_{m}^{2}+\nu _{l}^{2})}.
\label{eq:Fw}
\end{equation}
Hereafter, instead of $\omega_m$ we assume its discretized version 
$\widetilde{\omega}_m$, and in the following  the tilde will be omitted.

The expression (\ref{eq:corrAA}) is the well known Wick rotated 
Green function (in the frequency representation) for a single free particle 
in the phonon field theory, which has in our case one spatial dimension.

In the quasiclassical regime the amplitudes of phonon oscillations are small and 
the interactions between phonons due to anharmonicity of the Hamiltonian (\ref{H_FK}) 
are negligible. At higher $\hbar$, the amplitudes of phonon vibrations grow 
as $\hbar^{1/2}$ and their interactions become more important. In general, 
interactions can essentially modify the Green function (\ref{eq:corrAA}) for 
phonon excitations.  This actually happens for $\hbar>\hbar_c$ where the spectrum
of excitations is significantly changed.
However, one may expect that the  spectral function of elementary excitation 
remains of the same form 
\begin{equation}
  F^{(l)}(\omega _{m})=
  f\frac{\hbar }{(\omega_{m}^{2}+\nu _{l}^{2})},
\label{eq:Fw_renorm}
\end{equation}
which differs from (\ref{eq:corrAA}) by the renormalized frequency value $\nu _{l}$
and by an overall renormalization factor $f$, in analogy with the Green function 
behavior in renormalizable quantum field theories (see, e.g. \cite{Shur98}). 
In fact, this idea is well supported by numerical data.

An extended elementary excitation involves all particles in the chain. In turn,
the Fourier harmonics of any particle coordinate in the chain are a
sum of contributions of many elementary excitations
\begin{equation}
 \left\langle \left|a_{i}(\omega _{m})\right|^{2}\right\rangle =
 \sum _{j}f^{(j)}_i\frac{\hbar }{(\omega_{m}^{2}+\nu _{j}^{2})},
\label{eq:Fw_x_i}
\end{equation}
where the sum goes over all chain excitations.
In the deep quasiclassical case the main contribution comes from phonons modes, 
and the sum goes over phonon modes $l$, with 
$f^{j}_i\to\left|V_{i}^{(l)}\right|^{2}$ and $\nu_{j}\to\nu_l$.

The goal of our study here is a complete picture of low-lying quantum
excitations in the chain, both for low and high values of $\hbar $.
In fact, the spectrum of low-lying excitations is crucially dependent
on $\hbar $: there are domains with a \emph{qualitatively} different
behavior. In order to illustrate this, let us consider the amplitude
of quantum motion of particles in the chain, given by Eq.(\ref{eq:Fw_x_i}).
It is seen that the contributions of low-frequency modes(small $\nu_l$) dominate in the
limit $\omega _{m}\rightarrow 0$, provided that their wave-function
profiles $V_{i}^{(l)}$ are not small at the particle position $i$.
Therefore the spectral function (\ref{eq:Fw_x_i}), computed in this limit,
gives a rough estimate for the frequency $\nu_l$ of the lowest mode. 

\begin{figure}[th]
\includegraphics[  clip,
  width=105mm,
  height=85mm,
  keepaspectratio,
  angle=270,
  origin=c]{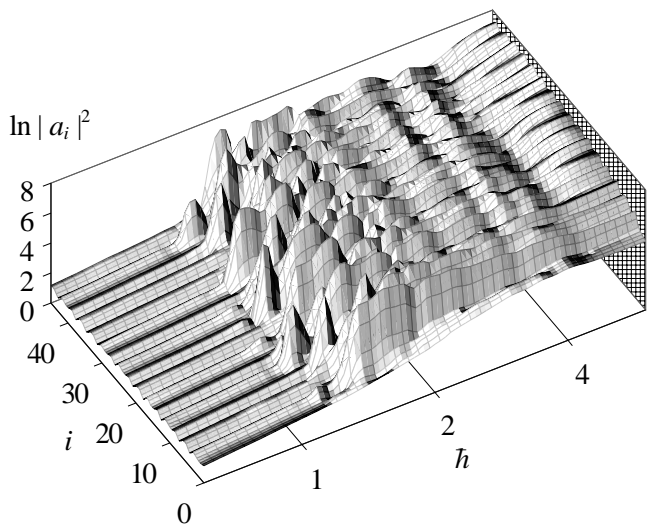}
\vskip -10mm
\caption{\label{fig:overview}Dependence of the amplitude of the lowest Fourier
harmonic $a_{i}\equiv a_{i}(\omega_1)$  on the particle
position $i$ in the chain at different $\hbar $. 
Here $\omega_1=\varpi =2\pi /\tau _{0}$, $m=1$. The chain parameters
are $s/r=89/55$, $K=5$, $\tau _{0}=80$. Typical number of iterations 
is $(1.5\div 5)\cdot 10^5$ at each
value of $\hbar$.}
\end{figure}

In Fig.\ref{fig:overview} the amplitude of the \emph{lowest} Fourier
harmonic $a_{i}(\omega_1)$ with $\omega_1=\varpi =2\pi /\tau _{0}$ is plotted
as a function of the particle position $i$ at different $\hbar =0.6-8$.
One can see that the whole interval of $\hbar $ splits naturally in
\emph{three} regions of qualitatively different behavior: (i) the
quasiclassical region $\hbar \lesssim 1$ where the amplitudes $a_i(\omega_1)$ of
the harmonics are very small and depend on the particle positions in some regular way;
(ii) the transition region $1\lesssim \hbar \lesssim 2$,
where this dependence is highly irregular and interactions between instantons and
phonons are important; and (iii) the region $\hbar \gtrsim 2$,
where this dependence becomes regular again and where, as we shall see below, 
a new phonon branch appears. 
Let us note, that at $\hbar<1$ the regular structure along the chain
is quasiperiodical, which reflect a fact that a classical chain is built of ``bricks'' 
of two principal sizes \cite{Zhir01a}. Above $\hbar\approx 2$ bricks are ``melted''
and chain properties become even more homogeneous along the chain. In the 
intermediate region irregular peaks come from different non-overlapping
instantons contributions, which as 
any tunneling effects are highly sensitive to small variations of potential barriers.
Below their contributions are exponentially small, and above, as we see further, they 
overlap and form new \textit{sliding} phase of the system.

In the following we analyze in detail these regions, corresponding to different 
intervals of $\hbar$.

\subsection{Quasiclassical region $\hbar \lesssim 1$}
\label{sub:Quasiclassical}

For $\hbar\lesssim 1$ the tunneling between different 
metastable classical configurations is negligible, and the particles mainly 
vibrate around some classical equilibrium positions. In this case the elementary
excitations are phonons, and the quantization of the chain is reduced
to the quantization of phonon modes, see e.g. \cite{Hu01}. 

To single out low energy excitations we use the following approach.
Of course, a particular excitation can be selected if the corresponding mode
$V_{j}^{(l)}$ is known, but in general this is not a trivial task.
However, for low lying excitations one may expect that the modes have a simple harmonic
form\begin{eqnarray}
V_{j}(k_{l})\equiv V_{j}^{(l)} & = & \sqrt{\frac{2}{L}}\sin (k_{l}j),\label{eq:profV}\\
 &  & k_{l}=\pi l/L,\; (l=1,2,\ldots )\nonumber 
\end{eqnarray}
where the wavelength $\lambda _{l}\equiv 2\pi /k_{l}$ is
much larger than a characteristic size of inhomogeneity in the chain.
Direct numerical computations \cite{PhonFK} of phonon modes in the classical 
FK chain support this ansatz (\ref{eq:profV}).

\begin{figure}[ht]
\includegraphics[  clip,
  width=115mm,
  height=85mm,
  keepaspectratio,
  angle=90,
  origin=c]{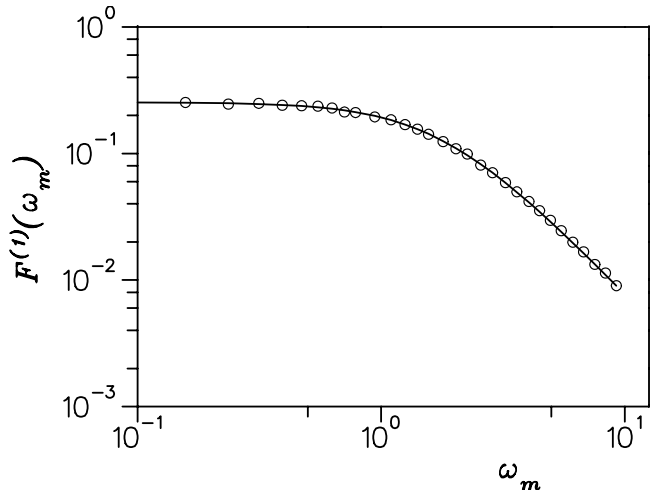}

\vskip -12mm
\caption{\label{fig:h08exmpla} The phonon spectral function $F^{(1)}(\omega _{m})$
versus the rescaled frequency $\omega _{m}$, data are shown
for the lowest spatial mode with $l=1$. The chain parameters are: 
$s/r=89/55$, $K=5$,$\tau _{0}=80$, $\hbar=0.8$. Here $m$ varies from $1$ to $89$,
but for clarity only selected values are shown. The solid
curve gives the fit by  Eqs.(\ref{eq:Fw},\ref{eq:Fw_renorm}), open circles show numerical
data. The fit determines the phonon frequency of the first mode ($\nu_1^2=3.170\pm0.012$)
and the renormalization factor $f=1.0034\pm0.0034$.}
\end{figure}

The numerical test of this anzatz (\ref{eq:profV}) is given in Fig.\ref{fig:h08exmpla}.
Here a typical result of quantum simulations of the spectral function
$F^{(l)}(\omega _{m})$ is shown for the lowest phonon
mode $l=1$ at $\hbar=0.8$. Fitting the data by Eq.(\ref{eq:Fw_renorm})
with the renormalization factor $f$ and the frequency of the
phonon mode $\nu _{l}$ as  free parameters, we obtain
$f=1.0034\pm 0.0034$ and $\nu_1^2=3.170\pm0.012$.
This fit shows that the renormalization factor $f$ is remarkably close
to unity, in spite of the fact that the value $\hbar =0.8$ is not
small. Hence, the ansatz (\ref{eq:profV})
provides a good approximation to actual profiles of lowest phonon
modes. This also indicates that the Gauss approximation used
in the Section \ref{sub:SpFunc} and in papers \cite{Hu98,Hu01r,Ho01}
works fine here. However we note that  at the same time the  quantum effects
renormalize substantially the phonon frequency ($\nu_1^2=3.170$)
compared to its value at $\hbar =0.2$  ($\nu _{1}^{2}=3.706\pm 0.019$).
The computations for the classical FK chain give $\nu _{1}^{2}=3.717$ ($\hbar=0$).

These data show that at small $\hbar$ the frequency $\nu_1$ obtained from
the quantum simulations approaches to the frequency of phonon mode in the classical chain.
This fact gives a very important check of the consistency of our quantum simulations. 
We stress that a good agreement between the numerical data 
of Fig.\ref{fig:h08exmpla} and the theoretical  spectral function (\ref{eq:Fw_renorm}) 
takes place in the \emph{whole} frequency range of $\omega _{m}$. 
This is, in fact, a very important consistency
check of the good relaxation of our paths ensemble at all frequencies,
including paths fluctuations at the lowest frequency available in
the system.

\begin{figure}[ht]
\includegraphics[ clip,
  width=115mm,
  height=80mm,
  keepaspectratio,
  angle=90,
  origin=c]{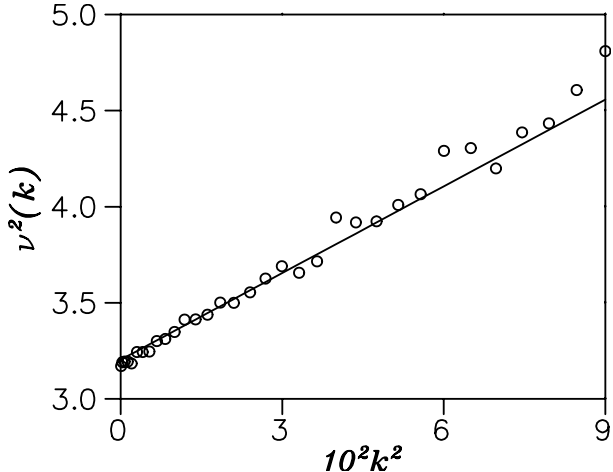}
\vskip -10mm
\caption{\label{fig:h08exmplb} The phonon dispersion law $\nu (k)$: open circles show 
the data obtained from the fit as it is shown in Fig.\ref{fig:h08exmpla} for $l=1\div30$,
and the straight line shows the fit given by Eq.(\ref{eq:DispLaw}). The chain
parameters are the same as in Fig.\ref{fig:h08exmpla}, the wave vector $k=\pi l/L$.}
\end{figure}

By fitting the data for different phonon modes $l=1\div30$ one
can extract the dispersion relation for phonons  $\nu(k)$ 
where $k=\pi l/L$ (see Fig.\ref{fig:h08exmplb}). The majority of data
points follow the straight line given by the formula
\begin{equation}
\nu ^{2}(k)=\nu _{0}^{2}+c^{2}k^{2}
\label{eq:DispLaw}
\end{equation}
where $\nu _{0}$ is the phonon frequency gap, and $c$ is the velocity
of sound. The fit of numerical data gives: $\nu_0^2=3.204\pm 0.026$,
$c^2=15.0\pm 0.7$ for $\hbar=0.8$; $\nu_0^2=3.706\pm 0.019$,
$c^2=13.3\pm 0.5$ for $\hbar=0.2$. These quantum data should be compared
with the classical case where $\nu_0^2=3.697$, $c^2=11.5$ ($\hbar=0$).
We note that there is a difference between the frequency of the first
spatial harmonic $\nu_1$ and the frequency gap value $\nu_0$ obtained
from the dispersion law. However this difference is small and comparable
with the statistical errors. For small $\hbar$ the parameters of the 
dispersion law converge to their classical values. 

The described approach allows to obtain a complete information about 
low-energy phonon excitations in the whole quasiclassical region 
$\hbar \lesssim 1$.

\subsection{Transition region $1\lesssim \hbar \lesssim 2$.}
\label{sub:Transition-region}

As it is seen from Fig.\ref{fig:overview}, this region corresponds
to the transition between two regimes $\hbar \lesssim 1$ and 
$\hbar \gtrsim 2$ where the dependence of the quantum excitations on the particle
location in the chain looks quite regular.

As it will be shown later the irregular behavior in the region
$1\lesssim \hbar \lesssim 2$ is related to a significant increase
of the density of instantons. At high density the interaction between
instantons becomes important and results in onset of new phonon branch 
at $\hbar>2$.
In this section we discuss the properties of instantons and phonons 
and obtain estimates for their frequencies. 

Let us start with a discussion of tunneling effects. For $\hbar < 1$
the contribution of tunneling to the spectral function is exponentially
small being proportional to $\exp (-const/\hbar )$. However at  $\hbar >1$ 
the tunneling probability becomes large and it gives a significant contribution
to the spectral function. The transition to this regime is seen in  Fig.\ref{fig:overview} 
as a sequence of sharp isolated peaks. 
Following the pioneering paper \cite{Poly77}, a tunneling event can
be associated to an \emph{instanton}. In the imaginary time representation,
the instanton is a local jump between two wells, which is fast compared
to the mean time interval between subsequent jumps: while the size
of instantons (in time $\tau$) is practically independent of $\hbar $, the
separation between them is exponentially large in the quasiclassical
limit (low instanton density). In particular, this means that in the 
first approximation one may consider instantons as independent jumps, 
as can be also checked from a direct examination of our path ensemble.
Here we should stress two important properties of instantons
in the quantum FK chain:

(i) each jump of a particle $i$ in its 
position $x_i$ gives displacements of neighboring particles,
which decay exponentially with the distance from the jump
location, i.e. instantons are exponentially localized in space inside 
the chain;

(ii) instantons are distributed inhomogeneously along the chain since
the tunneling probability is highly sensitive to variations of barrier
heights due to chain inhomogeneity. 

A simple explanation of the exponential localization of an instanton
(along the chain) comes from the fact that the new static configuration
produced by it can be seen as a local \emph{static} defect on the
original configuration, which is known to dye away exponentially with
the classical Lyapunov exponent \cite{Au78,Au83b,Au83c}. Indeed,
in Fig.\ref{fig:overview} one can see that at $\hbar \sim 1.2\div 1.3$
instanton contributions are exponentially peaked around some particular
positions along the chain. 

Let us now consider the properties of elementary excitations originated
by instantons. There is a question how to select numerically a single instanton 
excitation.  Obviously, the ansatz (\ref{eq:profV}) used for phonons is good 
only for extended modes while instantons are localized in space. Therefore 
in this case we analyze numerically the frequency spectrum of a given
particle $i$ in the chain (defined by Eq.(\ref{eq:four_x})). If
$\hbar $ is not too high (close to one), instantons do not overlap, and the
main contribution to the spectrum comes from the instanton which is near to the given 
particle. This contribution reaches its maximum for a particle which actually
jumps.

At the same time besides instanton jumps the quantum motion of a particle 
in the chain contains a contribution of many phonons with different frequencies 
(see, e.g. Eq.(\ref{eq:Fw_x_i})). This phonon background should be subtracted in 
order to single out the contribution of instanton.  
Fortunately the frequencies $\nu _{l}$ of phonon excitations are much higher than 
the frequency $\nu ^{(inst)}$ of chosen instanton.
Hence these two types of excitations are well separated in the frequency domain.
Therefore in our analysis of the instanton contribution to the spectral function 
$F_i(\omega_m)\equiv\langle \left| a_{i}(\omega _{m})\right|^2\rangle$ 
we can restrict ourselves to the frequency domain $\omega _{m}<\omega _{bound}$,
where the boundary $\omega _{bound}$ is chosen by the condition 
$\nu ^{(inst)}\ll \omega _{bound}\ll \nu _{l}$
for all phonon modes $l$. In this frequency domain the phonon contribution
(\ref{eq:Fw}) is practically independent of $\omega _{m}$ and
can be replaced by some constant $C_{i}$. Thus we can extract
the frequency $\nu ^{(inst)}$ and the weight $f_{i}^{(inst)}$ of
the instanton excitation by following fit for the spectral function 
$F_{i}(\omega _{m})$:

\begin{equation}
 F_{i}(\omega _{m})=f_{i}^{(inst)}\hbar /\left(\omega _{m}^{2}+
 (\nu _{i}^{(inst)})^{2}\right)+C_{i}.\label{eq:fitInst}
\end{equation}
This fit contains three free parameters $\nu _{i}^{(inst)}$, $f_{i}^{(inst)}$ and
$C_{i}$. 

\begin{figure}[ht]
\includegraphics[  width=115mm,
  height=85mm,
  keepaspectratio,
  angle=90,
  origin=c]{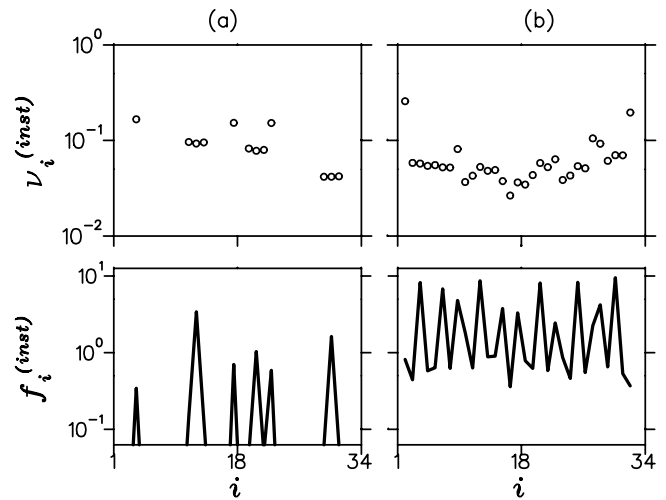}
\vskip-12mm
\caption{\label{fig:xEinst}Dependence of the instanton frequency 
$\nu _{i}^{(inst)}$ (top)
and its weight $f_{i}^{(inst)}$ (bottom) on the position $i$ inside the chain 
(see text for explanations). (a) $\hbar=1.2$. Instantons do not overlap: 
the three highest peaks involves a group of three particles each. 
(b) $\hbar=1.8$ Instantons overlap. Chain parameters are
$s/r=34/21$, $K=5$, $\tau _{0}=320$.}
\end{figure}

If instantons do not overlap then one may expect that
there are groups of particles which motion is dominated by a single
instanton. Inside each group the fit (\ref{eq:fitInst})
should give the same values for the frequencies $\nu _{i}^{(inst)}$, 
while the variation of weight $f_{i}^{(inst)}$ with $i$ 
determines the instanton profile along the
chain. This case is illustrated in Fig.\ref{fig:xEinst}a, which
corresponds to the early onset of instanton contribution at $\hbar =1.2$.
The six peaks in the bottom part of Fig.\ref{fig:xEinst}a show six non 
overlapping instantons, while the top part shows the corresponding frequencies 
as a function of particle index $i$ inside the chain. The peaks have different 
amplitudes, and the highest three of them involve groups of three particles
which have the same frequency inside each group. 

At higher $\hbar $
the number of instantons starts to grow rapidly and they begin to overlap.
A direct confirmation of this trend is seen in Fig.\ref{fig:xEinst}b
which corresponds to $\hbar =1.8$. Here all instantons have about ten percent
overlap with their neighbors and their interaction is rather
strong. As a result the step-like structure of frequencies, seen at
the top of the Fig.\ref{fig:xEinst}a, is practically destroyed. Thus
the instantons are ``collectivized'' and the consideration
of a single instanton as one-particle jump over barrier becomes 
not adequate. At higher $\hbar$ values this process leads to appearance
of a new phonon mode.  

\begin{figure}[th]
\includegraphics[  width=115mm,
  height=85mm, 
  keepaspectratio,
  angle=90,
  origin=c]{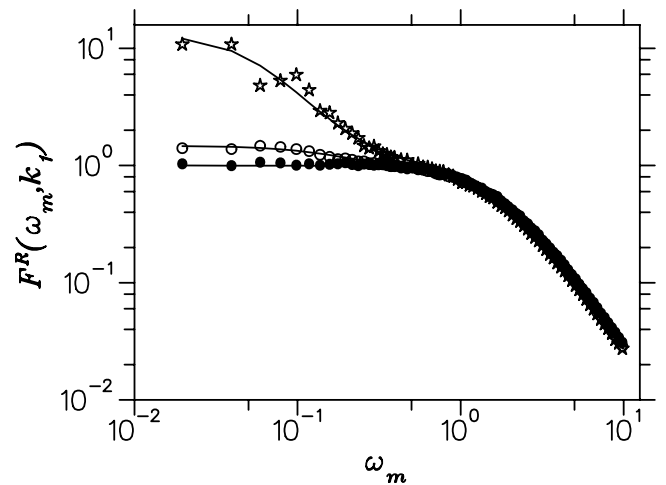}
\vskip -12mm
\caption{\label{fig:spIP}Rescaled spectral function 
$F^{R}(\omega ,k_{1})=(\nu _{1}^{2}/\hbar )F(\omega ,k_{1})$
at the very beginning of instanton onset. The bump which appears at 
$\omega\lesssim 0.2$ corresponds to instanton admixture to the phonon 
spectral function. Black points, open circles and stars
correspond to $\hbar=1,1.1$ and $1.2$.
Lines show the fit to numerical data with Eq.(\ref{eq:fitPhIn}). Chain parameters
used in simulations are: $s/r=34/21$, $K=5$,$\tau _{0}=320$.}
\end{figure}

Let us now discuss the phonon properties in the region
$1\lesssim \hbar \lesssim 2$. As in the previous Section \ref{sub:Quasiclassical},
we extract them from the the spectral
function $F(\omega _{m},k_l)\equiv F^{(l)}(\omega _{m})$ obtained on the basis 
of the ansatz (\ref{eq:profV}) for a phonon mode $l$.  
However, in this region of $\hbar $
the spectral function $F^{(l)}(\omega _{m})$ has an admixture of instantons,
which grows rapidly with $\hbar $: a change
of $\hbar $ from $1.1$ to $1.2$ results in more than ten times
of the admixture weight(see Fig.\ref{fig:spIP}). We note that 
in the absence of instanton contribution the rescaled 
phonon spectral function 
$F^{R}(\omega ,k_{l})\equiv(\nu _{l}^{2}/\hbar )F(\omega ,k_{l})$
plotted in Fig.\ref{fig:spIP} should have an
universal limit equal to unity independent of the value of $\hbar$.
Hence the increase of $F^{R}(\omega ,k_{l})$ at small $\omega$
stresses the important contribution of instantons.
These instantons have different frequencies and their contribution
to the spectral function can be rather complicated.  
However, we can use again the strong frequency separation between instanon 
and phonon excitations. Indeed, for $\omega _{m}\gtrsim \nu _{l}\gg \nu ^{(inst)}$, 
all instanton contributions have an universal behavior  
$\propto \omega _{m}^{-2}$. Therefore we may replace them by a single 
``instanton contribution'' with some average instanton frequency $\bar{\nu}^{inst}$. 
Then the spectral function can be fitted by a sum of two contributions:
\begin{eqnarray}
  F(\omega_{m},k_{l}) & = & f_{ph}\hbar /\left(\omega_{m}^{2}+
  \nu_l^{2}(k_{l})\right)\nonumber \\
   &  & +f_{inst}\hbar /\left(\omega_{m}^{2}+(\bar{\nu}^{inst}(k_{l})^2)\right)
\label{eq:fitPhIn}
\end{eqnarray}
where $f_{ph}(k),\nu_l(k)$ and $f_{inst}(k),\bar{\nu}^{in}(k)$ are
free fit parameters for phonons and instantons, respectively. 

\begin{figure}[th]
\includegraphics[  clip,
  width=110mm,
  height=85mm,
  keepaspectratio,
  angle=90,
  origin=c]{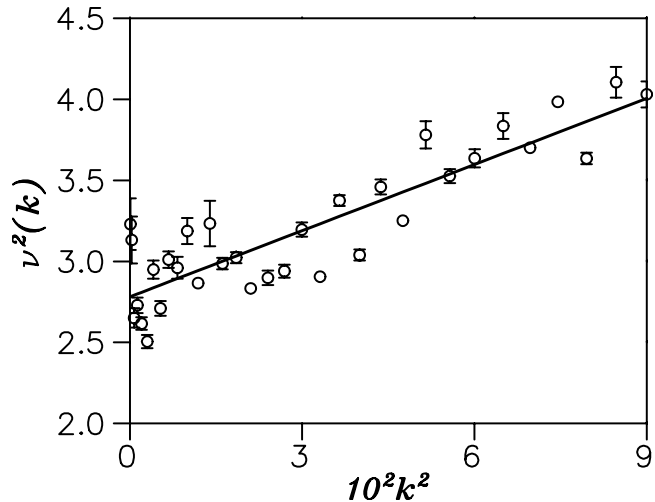}
\vskip -10mm
\caption{\label{fig:DLinTR} Frequency $\nu (k)$ of phonons
versus wavenumber $k$ obtained from the fit (\ref{eq:fitPhIn})
at $\hbar =1.5$ (in the middle of the transition region).
Chain parameters used in simulations
are: $s/r=89/55$, $K=5$,$\tau _{0}=80$.}
\end{figure}

Fitting the data for different phonon modes at $l=1-30$ we extract
the phonon dispersion law (\ref{eq:DispLaw}) (see Fig.\ref{fig:DLinTR},
compare with fit procedure for $\hbar<1$).
Contrary to the case of Fig.\ref{fig:h08exmplb} at $\hbar<1$, now
the data for the dispersion law $\nu (k)$ are scattered inside  
some finite band.
This indicates that the anzatz (\ref{eq:profV})
for the phonon profile is not so good to single out particular phonon
modes. Actually, the width of the
band provides some measure of the inaccuracy. Nevertheless, the phonon
modes are still approximately defined in the domain of $\hbar $ under consideration.
Their frequency decreases as $\hbar \to 2$ but remains separated
from zero by a finite gap.

We also note that the fit (\ref{eq:fitPhIn}) allows formally
to determine the dispersion law $\nu _{inst}(k)$ for the instanton branch.
However the numerical data give irregular scattering of points inside a band
$0 \leq \nu_{inst}^2 \lesssim 0.01$ without any clear dependence on $k$. 
%(see Fig.\ref{fig:DLinTR}b).
The reason of such behavior is simple: projections of irregular positions of instantons
on the harmonic ansatz (\ref{eq:profV}) produce random weights for
contributions of different instantons. This result represents another 
manifestation of glass-like structure formed by instantons frozen/pinned
inside the chain. Since the positions of instantons are random the phonons
cannot propagate along the chain on large distances. In fact they become
localized by disorder in a way similar to the one-dimensional Anderson localization
(more details on the phonon properties in this regime will be presented elsewhere
\cite{PhonFK}).

\subsection{New sliding phonon branch at $\hbar >2$}
\label{sub:New-phon}

\begin{figure}[ht]
\includegraphics[  width=115mm,
  height=85mm,
  keepaspectratio,
  angle=90,
  origin=c]{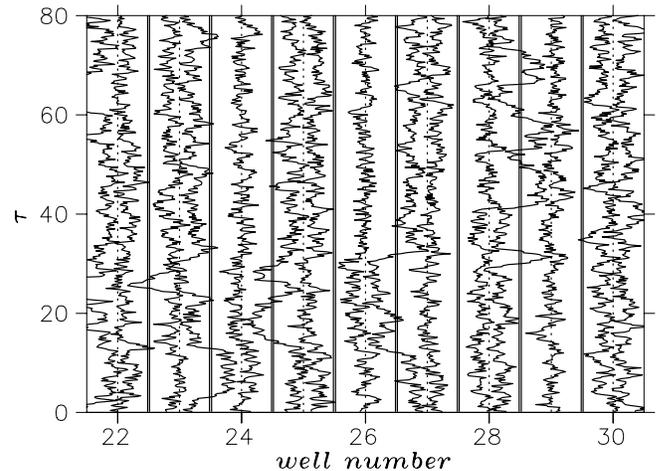}
\caption{\label{fig:path23} A sample of quantum paths of particles inside 
some chain fragment, which corresponds to periods of the external potential 
with numbers $22\div30$. Dashed lines show bottoms of the wells, thick solid
lines show the tops of the barriers.
Note an example of highly correlated instanton transitions at $\tau=15-30$
which involve particles  in up to 5 periods of the potential.
Parameters of the simulation are: $\hbar=2.3$, $s/r=89/55$, $K=5$ and 
$\tau _{0}=80$. }
\end{figure}

From Fig.\ref{fig:overview} one can see that the variation 
of the amplitude $a_{i}(\omega_1)$ with $i$ (low frequency excitations)
becomes rather smooth at $\hbar> 2$. 
This means that the instantons are strongly overlapped here.
In fact, this means 
that several particles of some chain fragment jumps from one well to the 
next simultaneously (see Fig.\ref{fig:path23}). 
This is nothing but sliding of a local chain fragment
along the periodical potential. The typical size of such fragments should
grow with $\hbar $. If their sizes reach the size of the chain  then the sliding
mode becomes open, and the phonon gap disappears. At this point the pinned
instanton glass turns into the sliding phonon gas.

\begin{figure}[th]
\includegraphics[  width=115mm,
  height=85mm,
  keepaspectratio,
  angle=90,
  origin=c]{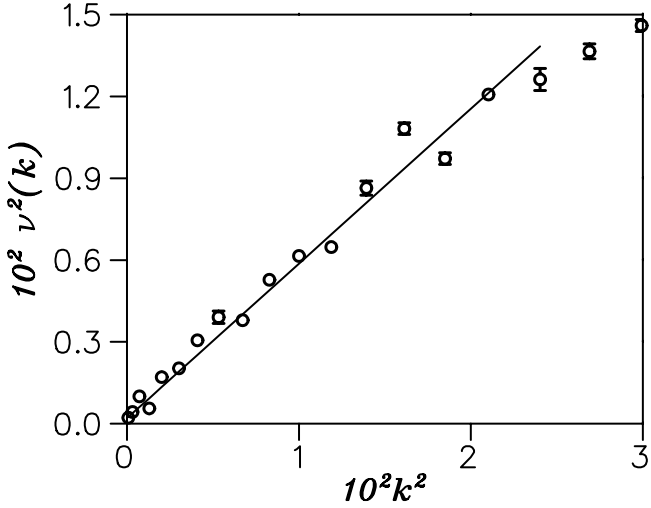}
\vskip -10mm
\caption{\label{fig:DLinSR} The dispersion law $\nu (k)$ for sliding 
phonons at $\hbar =2.5$. The numerical data (circles with error bars)
are obtained from the fit (\ref{eq:fitPhIn}) for the instanton branch
$\nu_{inst}(k_l)$.
The straight line shows the best fit (\ref{eq:DispLaw}) to numerical data. The chain
parameters used in simulations are: $s/r=89/55$, $K=5$,$\tau _{0}=80$.}
\end{figure}

A confirmation of this picture is presented in Fig.\ref{fig:DLinSR},
which corresponds to $\hbar =2.5$ being just above the transition point
$\hbar_c\approx 2$. The numerical data for the dispersion law $\nu (k)$
in Fig.\ref{fig:DLinSR} are obtained from the fit (\ref{eq:fitPhIn}) 
at $\hbar >2$. Here the behavior of phonon and instanton modes changes 
dramatically: phonons data $\nu (k)$ are now irregularly scattered over a wide 
band while the data points for instanton branch follow a single line, 
reproducing fairly well a \emph{phonon}-like dispersion law with  zero gap
(see Figs.\ref{fig:DLinSR},\ref{fig:Esurf}).
In particular, the fit (\ref{eq:DispLaw}) gives 
$\nu _{0}=0.04\pm 0.01$, which is close to zero. In fact, 
this value is smaller 
than the minimal frequency $2\pi /\tau _{0}$ ($\approx 0.079$, 
at $\tau _{0}=80$) and therefore it is compatible with zero. 

On the contrary, the wide scattering of data points for the phonon
branch indicates that the ansatz (\ref{eq:profV}) is not good for phonon 
contribution at $\hbar\gtrsim 2$ (see Fig.\ref{fig:Esurf}). This scattering of points
is related to the localization of high-frequency
phonon modes. In contrast, a smooth behavior of data points for the instanton
branch demonstrates that the instanton wave functions become close
to the harmonic wave ansatz (\ref{eq:profV}). Hence these excitations
are delocalized. This leads us to the conclusion, that 
for $\hbar>2$ the instanton branch is replaced by a new gapless
branch of new sliding phonons.

\subsection{Global picture of elementary excitations in the FK chain}
\label{sub:SummaryExcit}

\begin{figure}[ht]
\includegraphics[clip,
  width=120mm,
  height=88mm,
  keepaspectratio,
  angle=270,
  origin=c]{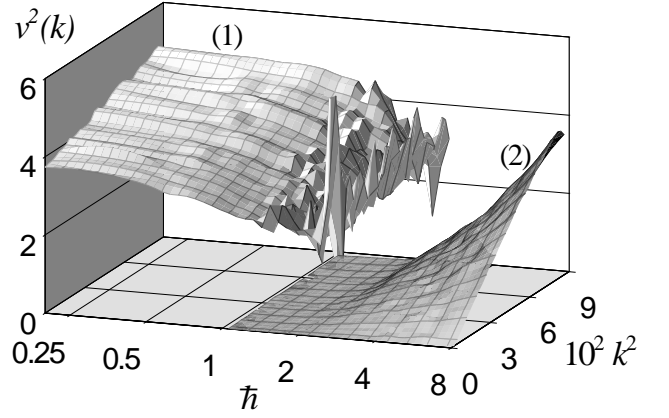}
\vskip -15mm
\caption{\label{fig:Esurf} The frequency $\nu (k)$ of elementary excitations
versus the wavenumber $k$ at different $\hbar $. Chain parameters used
in simulations are: $s/r=89/55$, $K=5$,$\tau _{0}=80$. The two sheets (1) and (2)
refer to phonon and instanton excitations respectively.} 
\end{figure}

The ensemble of data for the dispersion law $\nu(k)$ of elementary excitations
at different values of $\hbar$ is shown in Fig.\ref{fig:Esurf} by the two sheets
representing the phonon branch (1) and the instanton branch (2). 
The numerical data are obtained with the ansatz (\ref{eq:profV}) by the fit 
(\ref{eq:fitPhIn}). 

The sheet (1) refers to phonons originated from classical phonon
modes which are well reproduced in the limit $\hbar \to 0$. The frequencies
of these modes are well separated from zero by a large gap.  Therefore
one may say that  they  form the optical phonon branch. At
$\hbar \gtrsim 1$ these modes show a tendency to become softer, and at 
$\hbar \gtrsim 1.5$ their dependence on $k$ and $\hbar $ becomes irregular.
As it was explained in the previous section this irregular dependence 
is related to the glass-like structure of randomly pinned instantons
which density increases with the growth of $\hbar$.
The sheet (2) appears at $\hbar <2$ from the instanton contributions
into the Feynman path integral. For $\hbar>\hbar_c\approx 2$
this instanton branch turns into a new gapless branch of sliding phonons
(see the discussion related to Fig.\ref{fig:DLinSR}).

\begin{figure}[h]
\includegraphics[  width=115mm,
  height=85mm,
  keepaspectratio,
  angle=90,
  origin=c]{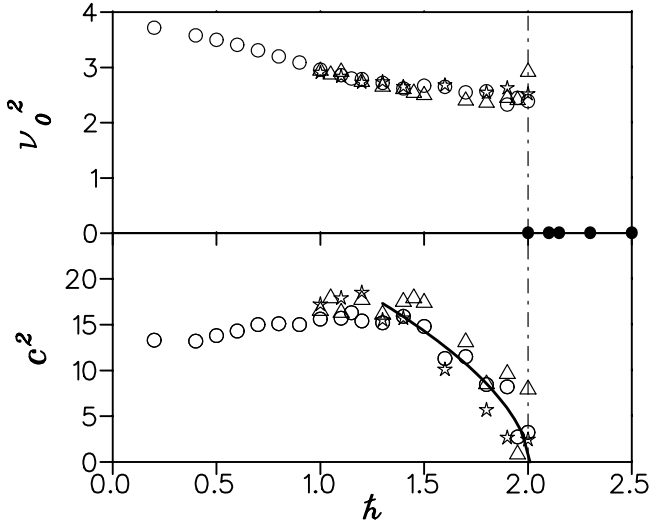}

\caption{\label{fig:PhSummary} Dependence of the phonon gap $\nu _{0}$ 
and sound 
velocity $c$ on $\hbar $ for the case of Fig.\ref{fig:Esurf} at $K=5$.
Open symbols correspond to the sheet (1) in Fig.\ref{fig:Esurf}.
Circles and triangles are obtained at $\tau_0=80$ for chain sizes 
$s/r=89/55$ and $233/144$ respectively. Stars correspond
to $s/r=34/21$ and $\tau _{0}=320$. The solid
line (bottom) gives the fit (\ref{eq:CritExpC2}) for  $1.3 \leq\hbar\leq 2$
(see text). The full circles (top) for $\hbar>2$ refer to the sliding 
phonon branch from the sheet(2) of Fig.\ref{fig:Esurf}, they 
indicate a zero phonon gap.}
\end{figure}

A more quantitative picture can be obtained from the
numerical data for the gap $\nu _{0}$ and the sound velocity $c$ for 
both sheets in Fig.\ref{fig:Esurf}.
For the optical phonon branch the values of $\nu_0$ and $c$ are determined
from the fit (\ref{eq:fitPhIn}) for different values of $\hbar$ 
(see Fig.\ref{fig:PhSummary}).  
The data show that the phonon gap remains finite and 
large at $\hbar<\hbar_c\approx 2$. 
In contrast, the sound velocity $c$ drops to zero as $\hbar$ 
approaches the value $\hbar_c$. This decay is compatible with
the fit 
\begin{equation}
  c^{2}=a(\hbar _{c}-\hbar )^{\alpha }
\label{eq:CritExpC2}
\end{equation}
shown by the solid line with $a=20.5\pm 1.3$, $\hbar _{c}=2.0\pm 0.1$
and the critical exponent $\alpha =0.52\pm 0.07$.
Even if the numerical data for $c$ have certain fluctuations they still 
clearly indicate the quantum phase transition at $\hbar_c\approx 2$,
where the sound velocity $c$ drops to zero. Further extensive numerical studies 
are required to determine the behavior in the vicinity of transition
in a more precise way.
\begin{figure}[h]
\includegraphics[  width=115mm,
  height=85mm,
  keepaspectratio,
  angle=90,
  origin=c]{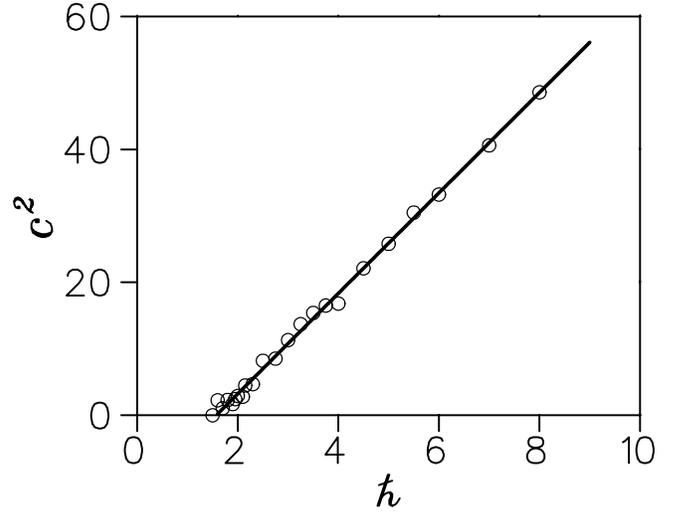}
\caption{\label{fig:InSummary} Dependence of sound velocity $c$
on $\hbar $ for excitations on the soft phonon sheet (2) 
in Fig.\ref{fig:Esurf}. Points correspond to chain parameters $s/r=89/55$ 
$K=5$ and $\tau _{0}=80$. The solid line shows the linear fit to the
sound velocity data inside the region $\hbar >2$ (see text).}
\end{figure}

For the sliding phonon branch  we extract the parameters $\nu_0$ and $c$ from the data of 
sheet (2) in Fig.\ref{fig:Esurf} using a more general fit 
given by 
\[
  \nu ^{2}(k)=\nu _{0}^{2}+c^{2}k^{2}/(1+k^{2}/k_{B}^{2}).
\]
Compared to the standard case (\ref{eq:DispLaw}) 
 we introduce an additional parameter $k_B$
to take into account the saturation of $\nu(k)$ at large $k$ (see, 
Fig.\ref{fig:DLinSR},\ref{fig:Esurf}).
The numerical data show that the gap 
$\nu _{0}$ is small and does not exceed the minimal frequency in 
the system $\varpi =2\pi /\tau _{0}$. Hence, the gap $\nu_0$ is compatible with zero
(see Fig.\ref{fig:PhSummary} (top)). 
At the same time  
the sound velocity $c$ for the sliding phonon branch grows approximately
linearly with $\hbar $ (see Fig.\ref{fig:InSummary}). 
The best fit of numerical data for $\hbar>2$ gives 
\begin{equation}
  c^{2}=a\cdot (\hbar -b),\label{eq:C2ofHbar}
\end{equation}
with $a=7.6\pm 0.1$, $b=1.57\pm 0.05$. Formally, the value of $b$ is different
from the value $\hbar_c=2.0$ in (\ref{eq:CritExpC2}). However in view of large statistical
fluctuations both fits for optical phonons and sliding phonons are compatible with the 
quantum phase transition at $\hbar_c\approx 2$.

\subsection{Time correlations}
\label{sub:Time-corr}

Time correlations are closely related to the frequency spectrum of
elementary excitations in the system. Their analysis is probably the
most traditional way to extract properties of the elementary excitations,
see, e.g. \cite{Creu81,Shur84}. Below we discuss the connection
of this traditional method with 
our approach based on the Fourier spectrum of Feynman paths
described above. We compare the results obtained by these two different methods.

The one-particle time correlators are directly related to the Fourier spectrum
of Feynman paths, and are given by the following expression:
\[
 \left\langle x_{i}(0)x_{i}(\tau )\right\rangle =
 \frac{1}{\sqrt{\tau _{0}}}\sum _{\omega _{m}}
 \left|a_{i}(\omega _{m})\right|^{2}\exp \left(-i\omega _{m}\tau \right)
\]
In fact, many elementary excitations with different frequencies contribute 
to a particle motion in the chain. In order to
single out a particular phonon mode we study the correlators of
normal modes $X^{l}(\tau )=\sum _{i}V_{i}^{(l)}x_{i}(\tau )$
where $V_i^{(l)}$ are eigenvectors defined in Eqs.(\ref{eq:Om2})-(\ref{eq:Om3}).
Then from Eq.(\ref{eq:corrAA}) we obtain
\begin{eqnarray}
 \left\langle X^{l}(0)X^{l}(\tau )\right\rangle  &  & =
 \frac{1}{\sqrt{\tau _{0}}}\sum _{\omega _{m}}\left|A^{l}(\omega _{m})\right|^{2}
 \exp \left(-i\omega _{m}\tau \right)\nonumber \\
 = &  & \left\langle \left(X^{l}(0)\right)^{2}\right\rangle 
 \left(\textrm{e}^{-\nu _{l}\tau }+
 \textrm{e}^{-\nu _{l}(\tau _{0}-\tau)}\right),
\label{eq:CorrPhT}
\end{eqnarray}
where $\left\langle \left(X^{l}(0)\right)^{2}\right\rangle =\hbar /2\nu _{l}$
is the contribution of a single phonon mode and the periodicity along
the time torus results in a second exponential term in (\ref{eq:CorrPhT}). 

However, Eq.(\ref{eq:corrAA}) assumes only small 
quantum fluctuations ($\sim \hbar $) around some \emph{classical} trajectory. 
But due to tunneling effects (or instantons) a particle jumps from one
\emph{classical} trajectory to another and its actual motion
is given by a sum of phonon $x_{i}^{(ph)}(\tau )$
and instanton $x_{i}^{(inst)}(\tau )$ contributions:
\[
  x_{i}(\tau )=x_{i}^{(ph)}(\tau )+x_{i}^{(inst)}(\tau ).
\]
In general, both motions influence each other, but in the quasiclassical
limit $\hbar \to 0$ they can be considered as independent. In this limit they
have quite different frequency scales: the phonon
frequency $\nu_l$ is of the order of $K^{1/2}$ while the frequency of tunneling jumps
$\nu_{inst}$
is exponentially small. In contrast, while the amplitude of phonon
oscillations in the limit $\hbar \to 0$ is small: 
$\left\langle \left(x_{i}^{(ph)}(\tau )\right)^{2}\right\rangle \propto \hbar $,
the amplitude of jumps is defined by difference between equilibrium
particle positions in two neighbor wells (in our case it is $\sim 3\div 4$),
i.e. it does not depend on $\hbar $. Hence, for a jumping particle
$\left\langle \left(x_{i}^{(inst)}(\tau )\right)^{2}\right\rangle \sim 10$
which is not small even in the quasiclassical limit $\hbar \to 0$. 
Therefore the instanton contribution to the time correlator 
has the form \cite{Shur84}
%\[
% \left\langle x_{i}(0)x_{i}(\tau )\right\rangle _{inst}=
% \left\langle \left(x_{i}^{(inst)}(\tau )\right)^{2}\right\rangle \times \\
% \left(\textrm{e}^{-\nu _{inst}\tau }+ 
% \textrm{e}^{-\nu _{inst}(\tau _{0}-\tau )}\right).
%\]
\begin{eqnarray}
 \left\langle x_{i}(0)x_{i}(\tau )\right\rangle _{inst}=
 \left\langle \left(x_{i}^{(inst)}(\tau )\right)^{2}\right\rangle \times \\
 \left(\textrm{e}^{-\nu _{inst}\tau }+ 
 \textrm{e}^{-\nu _{inst}(\tau _{0}-\tau )}\right). \nonumber
\end{eqnarray}
The pre-exponent factor is large for jumping particles even in the deep quasiclassical 
regime.

In fact, not any particle can easily jump from one well to another: different
classical trajectories have different actions and all jumps which
result in a large change of action $\Delta S\gtrsim \hbar $ are
inhibited. In particular, it is clearly seen in Fig.\ref{fig:overview}
that the number of instanton peaks is smaller at smaller $\hbar$ since
the contribution of transitions with large difference in action $\Delta S$
is suppressed.

\begin{figure}[h]
\includegraphics[clip,
  width=115mm,
  height=85mm,
  keepaspectratio,
  angle=90,
  origin=c]{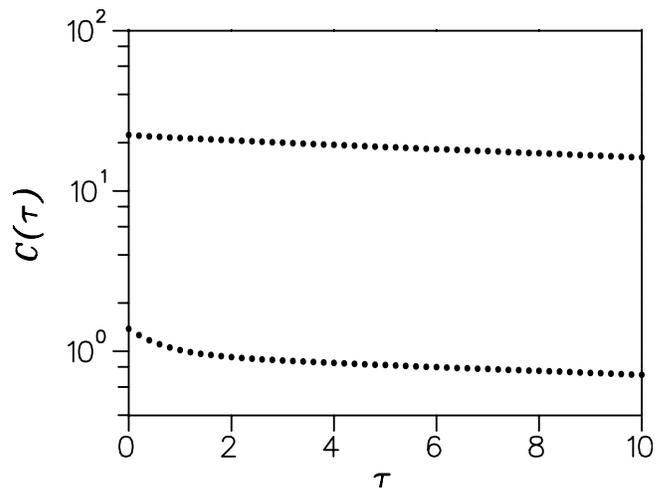}
\caption{\label{fig:CorrTa} The numerically computed time correlator 
$C(\tau )=\left\langle X^l(0)X^l(\tau )\right\rangle$ for $l=1$ at
different time separations $\tau $ for $\hbar =1.4$ (lower points) and 
$\hbar=2.5$ (upper points). The parameters of the chain are $s/r=89/55$ 
$K=5$ and $\tau _{0}=80$.
}
\end{figure}

Finally, the general form for the time correlator 
which takes into account the instantons contribution takes the form:
\begin{eqnarray}
 C(\tau ) & \equiv  & \left\langle X^{l}(0)X^{l}(\tau )\right\rangle =\nonumber \\
 & = & \left\langle \left(X^{l}(0)\right)^{2}\right\rangle 
 \left(\textrm{e}^{-\nu _{l}\tau }+\textrm{e}^{-\nu _{l}(\tau _{0}-\tau )}\right)
\nonumber\\
 & + & \sum _{inst}w_{in}\left(\textrm{e}^{-\nu _{inst}\tau }+
 \textrm{e}^{-\nu _{inst}(\tau _{0}-\tau )}\right).
\label{eq:CorrPhInT}
\end{eqnarray}
Here the first term describes the phonon contribution and in the second term
 the sum is taken over instantons and the instanton weights 
$w_{inst}=\sum _{i,k}V_{i}^{(l)}V_{k}^{(l)}
 \left\langle x_{i}(0)x_{k}(0)\right\rangle $
describe the overlap with the ansatz (\ref{eq:profV}). Both types 
of contributions (\ref{eq:CorrPhInT})
are clearly seen in Fig.\ref{fig:CorrTa} at $\hbar =1.4$. 
The initial rapid drop at $\tau\lesssim 1.5$ 
corresponds to phonon contribution while the slow decay
at $\tau>1.5$ corresponds to the instanton contribution. 
This initial drop is related to the existence of large
quasiclassical gap for phonon excitations.
For $\hbar=2.5>\hbar_c$ the gap disappears and the correlator decay
very slowly (see Fig.\ref{fig:CorrTa}).

\begin{figure}[h]
\includegraphics[clip,
  width=115mm,
  height=85mm,
  keepaspectratio,
  angle=90,
  origin=c]{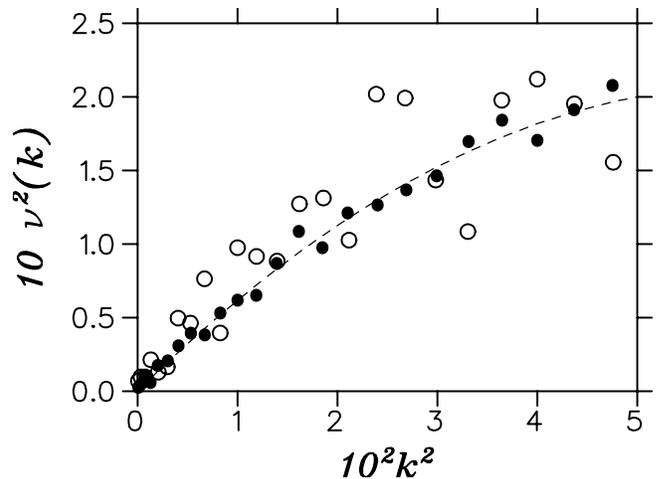}
\caption{\label{fig:CorrTb} Frequency of elementary excitations $\nu (k)$ 
versus $k$ for sliding phonon branch at $\hbar =2.5$. Open circles show 
data extracted from the  time correlator $C(\tau)$ for different $l$ 
(see Fig.\ref{fig:CorrTa}),
full circles present the results obtained from the fit (\ref{eq:fitPhIn})
for the frequency spectrum of Feynman paths (see Fig.\ref{fig:DLinSR}). 
The chain parameters are the same as in Fig.\ref{fig:CorrTa}.
}
\end{figure}

For $\hbar>\hbar_c$ we have a new phase where
instantons are replaced by sliding phonons.
Therefore in this regime we fit the numerical data for $C(\tau)$
by Eq.(\ref{eq:CorrPhInT}) with $w_{inst}=0$. The results for 
different $l$ allow to obtain numerically the dispersion law
$\nu(k)$ shown in Fig.\ref{fig:CorrTb} (open circles). However
the accuracy of this data is not so good compared to data
(full circles) obtained 
from the analysis of Fourier spectrum of Feynman paths described in the 
previous sections.
 
\section{quantum phase transition}
\label{sec:QFT}

The structual rearrangement of the elementary excitations spectrum can
be related to \emph{quantum phase transition} in the chain from a
``pinned'' to ``sliding'' phase. However, we would like to stress that
in contrast to the classical picture \cite{Au83b,Au83c} in the
quantum case the absence of energy gap for excitations is not 
necessarily related to the opening of the sliding phase. Indeed,
due to quantum tunneling through Peierls-Nabbarro barriers related to
instantons there are excitations
with energy which decreases exponentially with the increase of barrier heights. 
Formally, this corresponds to the disappearance of excitation gap. 
Therefore, to confirm firmly
the appearance of sliding phase we need to consider spatial correlations
of particle motion in the chain. The sliding phase appears when
the spatial correlation 
length becomes comparable with the length of the chain.

\subsection{Spatial correlation length}\label{sec:CorrLen}

The analysis of spatial correlations (correlations between the motion
of different particles in the chain) is the most evident way to observe
the transition between pinned and sliding phases. In principle, the
spatial
correlation function can be explicitly computed if the spectrum
and the wave functions of elementary excitations are known. In fact, we
have a complete quantitative picture for \emph{phonon} modes, at least
for the low-lying ones. To obtain numerically the value of the spatial
correlation length $l_c$ we assume that the elementary excitation
spectrum is given by the dispersion relation 
$\nu ^{2}(k)=\nu _{0}^{2}+c^{2}k^{2}$ 
and the corresponding phonon modes are given the ansatz 
(\ref{eq:profV}). Then the same-time spatial correlator reads:
\begin{eqnarray}
 \left\langle \left(x_{i}-\left\langle x_{i}\right\rangle \right)\left(x_{j}-
 \left\langle x_{j}\right\rangle \right)\right\rangle  & = & 
 \sum _{l}V_{i}^{(l)}V_{k}^{(l)}
 \left\langle \left(X^{l}(0)\right)^{2}\right\rangle \nonumber \\
 =\hbar \sum _{k}\frac{\cos (k(i-j))}{L\nu (k)} & = & 
  \frac{\hbar }{\pi c}K_{0}\left(\left|i-j\right|\right/l_c),
\label{eq:CorrX}
\end{eqnarray}
where $K_{0}(x)$ is McDonald's function, with a known asymptotics:
$K_{0}(x)\rightarrow _{x\rightarrow \infty }\sqrt{\frac{\pi }{2x}}\textrm{e}^{-x}$,
and $l_c= c/\nu _{0}$.

\begin{figure}[H]
\includegraphics[  width=65mm,
  height=85mm,
  angle=90,
  origin=c]{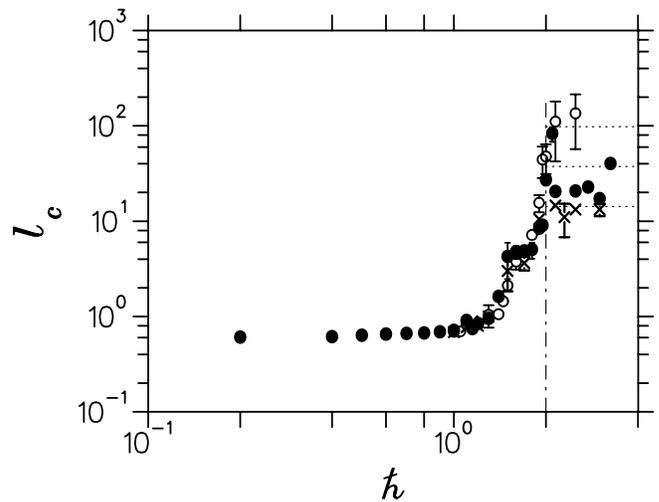}
\caption{\label{fig:CorrX}The dependence of the spatial correlation length 
$l_{c}$ on $\hbar $. Crosses,
full and open circles correspond to chains with $s/r=34/21$,
$89/55$ and $233/144$, respectively; $K=5$, $\tau _{0}=80$. 
The vertical dot-dashed line marks the quantum phase transition
at $\hbar_c\approx 2$.
The positions of the horizontal dotted lines are proportional to the chain length 
$L=2\pi r$.
}
\end{figure}

The fit (\ref{eq:CorrX}) of the numerical data gives the value
of $l_c$  for different values of $\hbar $ as it is
shown in Fig.\ref{fig:CorrX}. It is seen that the length $l_c$ has
a sharp increase at $\hbar_c \approx 2$, and for $\hbar>\hbar_c$
it becomes comparable with the length of the chain.
This indicates that we have the \emph{quantum phase transition}
near $\hbar_c\approx 2$.
Indeed at $\hbar <2$ the length $l_c$
is practically independent of the chain length $L$ while at $\hbar >2$
it starts to increase with $L$. This is confirmed by the data of
Fig.\ref{fig:CorrX} where in spite of strong fluctuations for $\hbar>2$
the length $l_c$ becomes comparable with the chain size $L$.
 
\subsection{Longwave response}

\begin{figure}[H]
\includegraphics[  width=115mm,
  height=85mm,
  keepaspectratio,
  angle=90,
  origin=c]{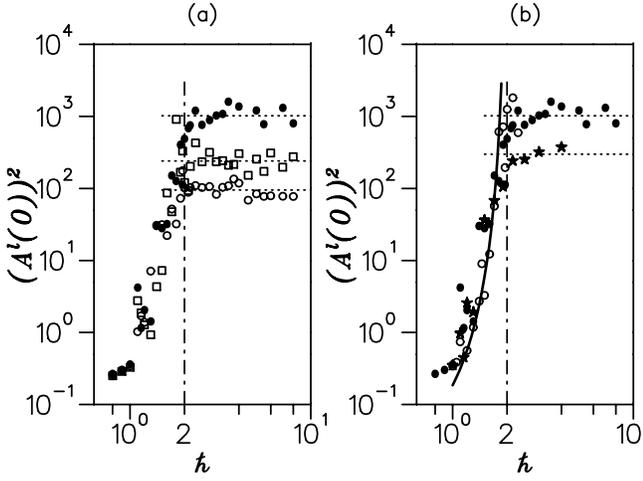}
\caption{\label{fig:X0sk}The amplitude square of zero-frequency quantum fluctuations
$(A^{l}(0))^2$ as a function of $\hbar$ for $K=5$ and $\tau _{0}=80$. 
(a) Full circles, squares
and open circles are for modes $l=1,3$ and $5$, respectively,
for the chain with $s/r=89/55$. (b) Open circles, full circles
and stars are for $s/r=233/144$, $89/55$ and $34/21$, respectively;
$l=1$. The vertical dot-dashed lines mark $\hbar_c\approx 2$,
the horizontal dotted lines give the average values for $\hbar>2$. 
}
\end{figure}

The appearance of the new sliding phonon phase implies
that the response of amplitude of the longwave modes should be large
in this regime. To test this expectation we present the dependence
of the amplitude $A^l(\omega=0)$ on $\hbar$ in Fig.\ref{fig:X0sk}.
The numerical data demonstrate a sharp increase of $A^l(0)$
near $\hbar_c\approx 2$. It is interesting to note that due to the existence
of frequency gap for phonon excitations at $\hbar<\hbar_c$ the amplitude $A^l(0)$
is not very sensitive to the variations of $l$. On the contrary, for 
$\hbar>\hbar_c$ the gap disappears and $A^l(0)$ starts to depend on $l$ 
(see Fig.\ref{fig:X0sk}a). In a similar way $A^l(0)$ is independent of
the chain length $L$ for $\hbar<\hbar_c$ while at $\hbar>\hbar_c$
it grows with $L$ (see Fig.\ref{fig:X0sk}b).
The numerical data of Fig.\ref{fig:X0sk}b for $1.5\leq\hbar\leq 2$
can be described by the fit
\begin{equation}\label{eq:gammaA2}
  \langle (A^{l}(0))^2\rangle\approx A (\hbar_c-\hbar)^{-\gamma} 
\end{equation} 
which gives $\gamma=5.06\pm 1.72$ and $\hbar_c=2.01\pm0.05$. 

\subsection{Other characteristics}

\begin{figure}[ht]
\includegraphics[  width=115mm,
  height=85mm,
  keepaspectratio,
  angle=90,
  origin=c]{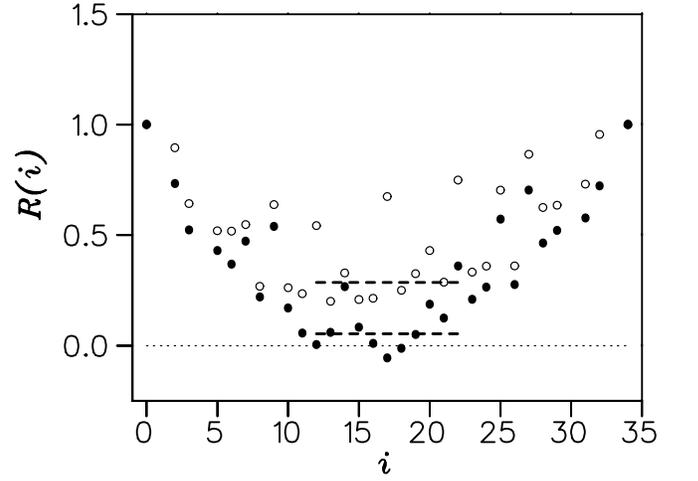}
\caption{\label{fig:slide} Dependence of the response function $R(i)$ on 
the particle position $i$
in the chain at $\hbar=1.8$ (full circles) and $\hbar=2.2$ (open circles).
The horizontal dashed 
lines show the interval of averaging for the minimal response value
$R_{min}$ in Fig.\ref{fig:CorrH}. Parameters of the chain
are $s/r=34/21$, $K=5$ and $\tau _{0}=80$. }
\end{figure}

Another way to test the transition from the pinned instanton glass to the sliding phonon phase
is to measure the sensitivity to small shifts of boundaries of the chain. With this aim we consider the shift of boundary particles $i=0$ and $i=s$ given by 
\begin{equation}
   x_0(\tau)=a_S \cos(2 \pi\tau/\tau_0), \;\; x_s(\tau)=x_0+L,
\end{equation} 
where the amplitude $a_S$ was fixed at $a_S=0.5$. In Fig.\ref{fig:slide} we present the
dependence of the response function 
$R(i)=\langle x_0 (x_i-\langle x_i\rangle\rangle/\langle x_0^2 \rangle$
on the particle number $i$ inside the chain.
It is seen that the response in the center of the chain 
drops strongly when the parameter $\hbar$ changes from 
$\hbar=2.2$  to $\hbar=1.8$. This means that the chain is locked for 
$\hbar<\hbar_c\approx 2$ while for $\hbar>\hbar_c$ the chain
slides following the displacements of the boundary
particles.

\begin{figure}[ht]
\includegraphics[  width=115mm,
  height=85mm,
  keepaspectratio,
  angle=90,
  origin=c]{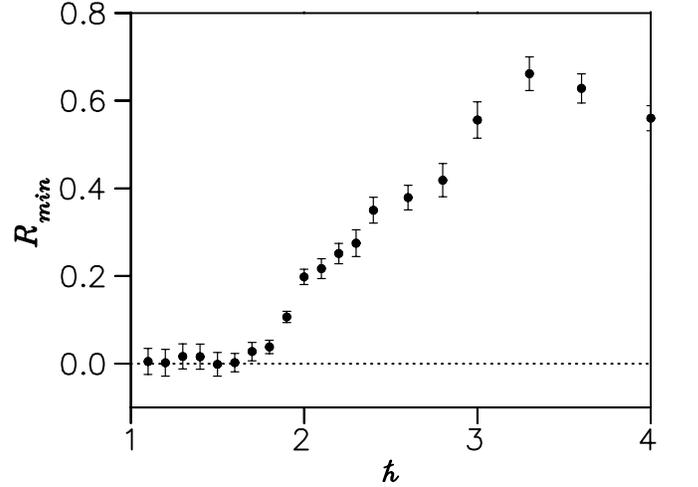}
\caption{\label{fig:CorrH} Dependence of the minimal response $R_{min}$ on $\hbar$
for the parameters of Fig.\ref{fig:slide}. 
}
\end{figure}

In order to get a more quantitative picture, we estimate the value of the response
function $R(i)$
at its minimum in the middle of the chain by taking its average value inside the central region at $i=12\div22$: 
$R_{min}=\langle R(i) \rangle_{i}$ (this interval is shown in Fig.\ref{fig:slide} by
horizontal dashed lines). The dependence of $R_{min}$ on the parameter $\hbar$ 
is presented in Fig.\ref{fig:CorrH}. We see, that the correlator in the central region 
of the chain deviates from zero at $\hbar>2$. 
According to the numerical data the response $R_{min}$ is very small for 
$\hbar<\hbar_c\approx 2$ while it becomes rather strong for $\hbar>\hbar_c$.
This confirms the existence of the quantum  phase transition 
from the pinned to sliding phase at $\hbar=\hbar_c\approx 2$.
\begin{figure}[ht]
\includegraphics[  width=115mm,
  height=85mm,
  keepaspectratio,
  angle=90,
  origin=c]{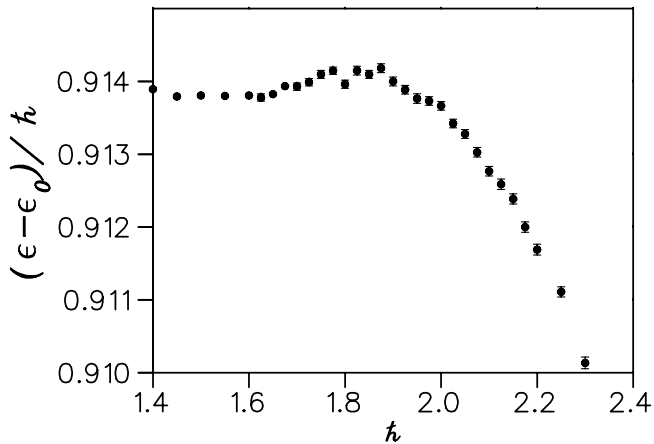}
\caption{\label{fig:DEvsH} Dependence of the total
 chain energy per particle 
on $\hbar$ for the chain parameters $K=5$, $s/r=89/55$ and $\tau_0=80$;
$\epsilon_0=5.302$ is an energy point chosen from convenience/illustration 
reasons.}
\end{figure}

What is the order of this transition? In our simulations we have repeated a cycle
changing slowly $\hbar$ from $\hbar=1$ 
to $\hbar=4$ and back about hundred times, but no hysteresis was found within
the statistical errors. Indeed,   
the difference of the chain energy per particle at the upward and backward paths 
did not exceed 
$10^{-3}$. This difference should be compared with the change of the energy per particle 
$\delta E\approx 2.4$ which takes place when $\hbar$ changes from $\hbar=1$ to $\hbar=4$.
The absence of hysteresis excludes  the phase transition of the first order. We 
also do not see any breaks in the dependence of the chain energy on $\hbar$. 
In order to make more visible small deviations 
from a linear law we plot in Fig.\ref{fig:DEvsH} the quantity 
$(\epsilon-\epsilon_0)/\hbar$, where $\epsilon_0=5.302$. The numerical data
show  that the slope of energy dependence changes near $\hbar = \hbar_c \approx 2$.
This change of slope is located  approximately at the same value of $\hbar$
where the divergence of the correlation length takes place (see  Fig.\ref{fig:CorrX}).
These data indicate that we have a second order quantum phase transition which appears
near $\hbar_c\approx 2$.
However, more extensive numerical simulations are required to determine more 
precisely the order of the transition.

\section{Conclusions}
\label{sec:Conclusions}

We have studied quantum tunneling phenomena in a particular model
of glassy material, the Frenkel-Kontorova chain. This system has
a lot of states, which are exponentially degenerate and (meta)stable
in the (quasi)classical limit. In the quantum case there are tunneling
transitions between these states that can be understood in  terms 
of instanton dynamics. In the quasiclassical limit at small $\hbar$ 
the instanton density is small and instantons are local and isolated.
With the increase of $\hbar $ their density grows and they start to overlap.

At sufficiently large $\hbar $ the instantons are coupled and
become  collectivized. As a result, the tunneling of particles 
in some fragments of the chain proceeds in correlated way.
The size of these correlations grows until it reaches the size of the system.
This leads to appearance of extended excitations
and opening of a new gapless sliding phonon branch.
Our data show that a quantum phase transition takes place
between the pinned and sliding phases. Absence of hysteresis effects,
as well as continuous dependence of chain energy on 
$\hbar$ exclude the first order phase transition, so we can classify
this transition as a continuous quantum phase transition. We stress
that the quantum phase transition from pinned to sliding phase
takes place in the regime where the classical chain always remains
in the pinned phase with the finite phonon gap.

The direct analysis of Fourier spectrum of Feynman paths ensemble allowed
to obtain detailed information on the dispersion law of low-lying 
excitations in both quantum phases. Nevertheless, some questions remain open for further
investigations. For example, one can analyze in more detail the effects 
of interactions between instantons at low density and study their
propagation properties in this regime. Another interesting remark 
concerns the behavior of the system in the vicinity of the transition
point at $\hbar_c\approx 2$. Indeed, in this region the kinetic
energy per particle is approximately 0.6 that is about ten times smaller
than the height of the potential barrier at $K=5$. Therefore more insights
are required to understand the underlying physics of this transition.
 
One can ask on how general are our results obtained in the frame 
of the Frenkel-Kontorova
model? In fact, we never used any specific features of this model
related to its nontrivial number theory properties. The only essential
point is the existence  of an exponential number of quasidegenerate 
states which is common for glassy materials and other disorder systems.
Therefore it is very interesting to study an analogous quantum
phase transition in systems with disorder and interactions.

This work was supported in part by the EC RTN network contract HPRN--CT-2000-0156;
OVZ thanks the groups in Como and Toulouse for their hospitality
during the work on this problem.
%\begin{acknowledgments}
%\end{acknowledgments}

%\bibliographystyle{revtex}
%%%\bibliography{fk}

\end{document}